\documentstyle[aas2pp4]{article}

\newcommand{\beq}{\begin{equation}}
\newcommand{\eeq}{\end{equation}}
\def\beqa{\begin{eqnarray}}
\def\eeqa{\end{eqnarray}}
\newcommand{\lsim}{\lesssim}

\def\onehalf{\textstyle {1\over 2}}
\def\onequarter{\textstyle {1\over 4}}
\def\p{\partial}

\begin{document}

\title{AN ISOCURVATURE CDM COSMOGONY. I. A WORKED EXAMPLE OF 
EVOLUTION THROUGH INFLATION}  

\author{P. J. E. Peebles}
\affil{Joseph Henry Laboratories, Princeton University,
Princeton, NJ 08544; pjep@pupgg.princeton.edu}
\authoremail{pjep@pupgg.princeton.edu}
 
\begin{abstract}

I present a specific worked example of evolution through inflation 
to the initial conditions for an isocurvature CDM model for
structure formation. The model invokes three scalar
fields, one that drives power law inflation, one that survives to
become the present-day CDM, and one that gives the CDM field a
mass that slowly decreases during inflation and so ``tilts'' the
primeval mass fluctuation spectrum of the CDM. The functional
forms for the potentials and the parameter values
that lead to an observationally acceptable model for structure 
formation do not seem to be out of line with current ideas about
the physics of the very early universe. I argue in an accompanying
paper that the model offers an acceptable fit to main
observational constraints.
\end{abstract}

\section{Introduction}

This paper with its companion (Peebles 1998; hereafter Paper II)
is the latest in a series of studies of isocurvature models
for structure formation. (Earlier papers may be traced back from
Peebles~1997). There are two motivations for this work. First, it
is important to know that there is an observationally viable and
theoretically not unreasonable alternative to the commonly
discussed adiabatic cold dark matter (ACDM) model for structure
formation. As long as there are alternatives it demonstrates that
we do not have an established standard model for the early
universe, that there is more to be discovered than tighter
constraints on parameters and functional forms for a potential.
Second, the isocurvature models seem to be better adapted to 
galaxy formation at high redshift, a condition I find attractive.
This paper is meant to demonstrate that initial conditions for
the isocurvature case, which I shall call ICDM, can be given an
explicit and not unreasonable basis in a physical model for
inflation. In Paper~II I argue that the model offers an
acceptable fit to the observational constraints that can be
applied without the use of numerical simulations.

There is considerable activity in the study of the rich
variety of ideas for a specific model for inflation 
motivated by current ideas in particle physics (as 
reviewed by Randall 1997). I make use of functional 
forms for potentials that commonly appear in these discussions, 
but the motivation is to obtain a specific model that could have
come out of inflation to compare to the rich suite of
observational evidence. The two cultures --- based in particle
physics and astronomical phenomenology --- will meet if
it becomes possible to establish that the narrow range of models
from fully acceptable physics overlaps the narrow range of models
that are observationally acceptable. 

The dynamical actors in the structure formation model to be
discussed here and in Paper~II are the same as in the family of
adiabatic CDM (ACDM) models --- baryons, radiation 
(the CBR) with initially homogeneous entropy per baryon, cold 
dark matter, and three families of neutrinos --- and in
about the same amounts. In the adiabatic version, a
scale-invariant departure from a homogeneous primeval mass
distribution has 
power spectrum $P\propto k$. In the ICDM model a  
scale-invariant spectrum of the distribution of the CDM would be 
$P\propto k^{-3}$, while the net mass density is homogeneous.
Early discussions of inflation accepted the proposition that
adiabatic and isocurvature scenarios are equally well motivated 
(eg. Steinhardt \& Turner 1983, Linde 1985; Seckel \&\ Turner
1985), but when Efstathiou \&\  
Bond (1986) demonstrated that a scale-invariant 
isocurvature model violates the bound on the CBR anisotropy
attention naturally turned to the scale-invariant adiabatic case.
The COBE detection of the CBR anisotropy (Smoot et al. 1992)
showed that if the universe were Einstein-de~Sitter the ACDM
spectrum with a reasonable bias would have  to be ``tilted'' from
scale-invariance, an arguably natural adjustment of the
inflation picture (eg. Crittenden {\it et al.} 1993). Tilt is not   
needed if the mean mass density (excluding a term in the
stress-energy tensor that acts like a cosmological constant) is
well below the Einstein-de~Sitter value, but the precedent has
been set: consider tilting the ICDM spectrum
to fit the observations. Models for inflation that
tilt the spectrum in the wanted direction
have been discussed by Kofman \&\ Pogosyan (1988), Salopek,
Bond, \&\ Bardeen (1989), and Linde \&\ Mukhanov (1997), 
and a model fitted to the observational constraints has been
presented in Peebles~(1997). Here and in Paper~II I present a
more detailed discussion along the lines of this last paper. 

For the purpose of displaying a specific worked example I adopt
definite values of the parameters in the cosmology and the
structure formation model and derive from them the 
parameters in the inflation model, the latter including the
needed initial conditions early in inflation. The parameters will
have to be reconsidered with each improvement of the
observations, of course. The hope is that such adjustments driven
by advances in the observations and physics may back us into the
corner of model and parameter space that is a reasonable
approximation to reality.  

The elements of the inflation scenario are presented in \S 2.
Section 3 starts from adopted values of the parameters in the
models for cosmology and structure formation and presents derived
values for the parameters in the inflation model. Except where
otherwise indicated units are chosen so $\hbar = 1=c$, and I
follow the notation in Peebles~(1993), \S 17. 

\section{The Inflation Model}

The model assumes three real single scalar fields, one whose
energy density drives homogeneous inflation, one that survives as
the present-day CDM, and one that serves to give the CDM field a
rolling mass that tilts the spectrum of quantum fluctuations 
frozen into the CDM distribution during inflation. A 
simple way to tilt to a near power law power spectrum of frozen
fluctuations starts with the condition that during inflation the
universe is expanding as a power of time (Abbott \&\ Wise, 1984; 
Lucchin \&\ Matarrese 1985),
\beq
	a\propto t^{1/\epsilon},\qquad
	H=\dot a/a =(\epsilon t)^{-1}\propto a^{-\epsilon}.
		\label{eq:a(t)}
\eeq
The past event horizon is eliminated (or rather can be very
large) if $\epsilon < 1$; a convenient value in the numerical
example below is $\epsilon =0.1$. 

The purpose of this discussion is to obtain a specific model for
the relevant properties of the cold dark matter. I do not address
the issue of how the energy density in the field that is driving
inflation is converted to the entropy that ends up mainly in the
CBR today.  

\subsection{Power Law Inflation}

Following Lucchin \&\ Matarrese (1985), 
the power law expansion law in equation~(\ref{eq:a(t)})
assumes the energy density during inflation is dominated by the
field $\chi (t)$ with potential energy density that is an
exponential function of the field, 
\beq
	V_\chi = {M^2\over\epsilon t_\ast ^2}e^{-\chi /M}.
		\label{eq:Vchi}
\eeq
Then the field equation is 
\beq
	{d^2\chi\over dt^2} + {3\over\epsilon t}{d\chi\over dt}
	= -{dV\over d\chi} = {M\over\epsilon t_\ast ^2}e^{-\chi /M},
\eeq
for the power law expansion in equation~(\ref{eq:a(t)}), 
and the adopted solution is
\beq
	e^{\chi /(2M)} = (t/t_\ast )(6 - 2\epsilon )^{-1/2}.
		\label{eq:chi}
\eeq
The energy density in the field is
\beq
	\rho  = {\dot\chi ^2\over 2} + V_\chi  = {6M^2\over\epsilon t^2},
		\label{eq:rho}
\eeq
where the dot means derivative with respect to proper time $t$.
Consistency of the Friedmann equation
\beq
	{\dot a^2\over a^2} = {8\over 3}\pi G\rho
\eeq
with equations~(\ref{eq:a(t)}) and~(\ref{eq:rho}) requires
\beq
	{M^2\over m_{\rm pl}{}^2} = {1\over 16\pi\epsilon},
		\label{eq:M}
\eeq
where $m_{\rm pl}=G^{-1/2}$ is the Planck mass.
Thus the mean mass density is
\beq
	\rho  = {3m_{\rm pl}^2\over 8\pi\epsilon ^2t^2}.
		\label{eq:rho1}
\eeq

Given $\epsilon$, equation~(\ref{eq:M}) fixes the mass $M$ in the
potential $V_\chi$ in equation~(\ref{eq:Vchi}) and sets the scale
for $\chi$. Then the constant $t_\ast$ fixes the time scale for
the evolution of $\chi$ as it passes through $\chi = 0$.

\subsection{The Cold Dark Matter}

If the Hubble parameter $H$ is close to constant during inflation
and the mass of the CDM field $\phi$ is much less than $H$ 
then the spectrum of fluctuations frozen into $\phi$ is the
scale-invariant form $P_\phi\propto k^{-3}$. To fit the
observations the fluctuations in $\phi$ must be tilted to favor
smaller scales (Peebles 1997).  
This is done by coupling $\phi$ to a field $\psi$ that gives
$\phi$ a mass comparable to $H$. The mass causes
decay of fluctuations of $\phi$ on scales larger 
the Hubble length, thus favoring the shorter
wavelengths that pass through the Hubble length later.

The Lagrangian density for the two fields is
\beq
	L = \onehalf\psi _{,i}\psi^{,i} 
		+ \onehalf\phi _{,i}\phi ^{,i} - V,
			\label{eq:L}
\eeq
where the potential energy density is
\beq
	V = \onequarter\beta\psi ^4 + \onehalf\mu ^2\psi ^2 + 
	\onehalf (\gamma \psi ^2 + m^2)\phi ^2. \label{eq:V}
\eeq
Here $\beta$ and $\gamma$ are dimensionless constants; in
the numerical example $\gamma\sim\beta\ll\epsilon$. As will be
described, the quartic term in $\psi$ causes the effective mass
of $\phi$ to roll to lower values at a rate that can be adjusted
to tilt the fluctuation spectrum of $\phi$  to an observationally 
acceptable power law, and the constant mass $\mu$ produces a
break in the power law to make $\langle\phi ^2\rangle$ converge. 
The constant mass $m$ is assumed to be much smaller than $H$
during inflation. After inflation, when $\psi$ has dropped to
zero, the CDM field has mass density  
\beq
	\rho ({\bf x}) = \onehalf m^2\phi ({\bf x})^2,\label{eq:rhophi}
\eeq
where $\phi ({\bf x})$ is the field frozen (or squeezed) during
inflation. This is the primeval mass distribution that seeds
structure formation in the ICDM model. 

The potential in equation~(\ref{eq:V}) is introduced {\it ad hoc}
to solve the problem of tilting the CDM fluctuation spectrum, 
but it will be argued in \S 4 that it does have a pedigree in
similar forms that appear in different contexts in many
discussion of inflation.  

It will be assumed that during inflation the field $\psi$ is
little affected by its coupling to $\phi$, $\psi$ is a
function of time alone, and $m\ll\psi$. Then the wave equations
from equations~(\ref{eq:L}) and~(\ref{eq:V}) are 
\beqa
	{d^2\psi\over dt^2} + {3\over\epsilon t}{d\psi\over dt}
	+ \beta\psi ^3 + \mu ^2\psi = 0,\nonumber\\
	{\p ^2\phi\over\p t^2} + {3\over\epsilon t}{\p\phi\over\p t}
	+\left( -{\nabla ^2\over a^2} +\gamma\psi ^2\right) \phi =0.
\eeqa
With the changes of variables
\beqa
	\tau = \epsilon ^{1/2}\mu t,\qquad
	\Psi = \beta ^{1/2}\mu ^{-1}\psi ,\nonumber\\
	a = a_1\tau ^{1/\epsilon },\quad
	\kappa = k/(\mu a_1), \label{eq:dim}
\eeqa
where $k$ is the comoving wave number of the Fourier mode
$\phi _k e^{i{\bf k}\cdot {\bf x}}$, the field equations are 
\beqa
	\epsilon {d^2\Psi\over d\tau ^2} + 
	{3\over\tau}{d\Psi\over d\tau} + \Psi +\Psi ^3 = 0,\nonumber\\
	\epsilon {d^2\phi _k\over d\tau ^2} +
	{3\over\tau}{d\phi _k\over d\tau} +
	\left( {\kappa ^2\over\tau ^{2/\epsilon }} +
	{\gamma\over\beta}\Psi ^2\right)\phi _k = 0.
		\label{eq:fieldeqs}
\eeqa
The adopted solution to the first equation at $\tau\ll 1$, where 
$\Psi\gg 1$, is 
\beq
	\Psi = {(3 - 2\epsilon )^{1/2}\over\tau },\quad
	\psi = \left( 3\epsilon ^{-1} - 2\over\beta\right) ^{1/2}
	{1\over t}. \label{eq:psiearly}
\eeq
Under the initial condition $\Psi\gg 1$ at $\tau\ll 1$ the field
$\Psi$ approaches this solution with increasing $\tau$. At
$\tau\gg 1$ the solution 
approaches a Bessel function, with amplitude that decays as 
\beq
	\Psi\sim\tau ^{-3/(2\epsilon )}.\label{eq:psilate}
\eeq

\subsection{The CDM Fluctuation Spectrum} 

The CDM field expressed as a quantum field operator is
\beq
	\hat\phi ({\bf x},t) = \sum {\hat a_{\bf k}\over (2kV_u)^{1/2}}
	\phi _k(t)e^{i{\bf k}\cdot {\bf x}} + {\rm hc},
		\label{eq:qft}
\eeq
where $\phi _k(t)$ is the solution to the second  of
equations~(\ref{eq:fieldeqs}). The normalization is fixed 
by the solution valid at small $\tau$ (and for $\epsilon < 1$),
where the frequency is large compared to the Hubble parameter 
$H$ (eq.~[\ref{eq:a(t)}]):
\beqa
	\phi _k \rightarrow a(t)^{-1}\exp -i\int ^\tau \omega d\tau,
		\nonumber\\
	\omega (\tau )^2= {1\over\epsilon}
	\left( {\kappa ^2\over\tau ^{2/\epsilon }} +
	{\gamma\over\beta}\Psi ^2\right) .\label{eq:bdy}
\eeqa
Equation~(\ref{eq:qft}) assumes space is periodic in some large
coordinate volume $V_u$, so the creation and annihilation
operators satisfy the discrete relation
$[\hat a_{\bf k},\hat a^\dagger _{{\bf k}'}]=\delta _{{\bf k},{\bf k}'}$.
With the normalization in equation~(\ref{eq:bdy}) the field momentum 
$\hat\pi =a^3\partial\hat\phi/\partial t $ satisfies the
canonical commutation relation 
$[\hat\phi ({\bf x},t),\hat\pi ({\bf x}',t)]=i\delta({\bf x}-{\bf x}')$.

With the initial condition in equation~(\ref{eq:bdy}), the mode
amplitude  
$\phi _k(t)$ approaches the constant value $\phi _k^o$ at large
$\tau$, and the field is frozen as a random Gaussian process with
autocorrelation function
\beq
	\xi _\phi (x_{12}) = 
	\langle\hat\phi ({\bf x_1})\hat\phi ({\bf x_1})\rangle ,
		\label{eq:xiphi}
\eeq
where the power spectrum is
\beq
	P_\phi (k) = \int d^3x\,\xi _\phi (x)
	e^{i{\bf k}\cdot {\bf x}} = |\phi _k^o|^2/(2k).
		\label{eq:pphi}
\eeq

Now let us estimate $\phi _k ^o$. The wavelength belonging to
comoving wavenumber $k$ passes through the Hubble length at
expansion parameter $a_k$, where 
\beq
	H_k=k/a_k,\qquad a_k\propto k^{1/(1 -\epsilon )}.
			\label{eq:crossing}
\eeq
The second expression follows from the expansion law in
equation~(\ref{eq:a(t)}). The characteristic epoch at which
$\Psi$ passes through unity is $\tau _1=1$
(eq.~[\ref{eq:psiearly}]), or
$t_1=\mu ^{-1}\epsilon ^{-1/2}$, when $H_1 = 1/(\epsilon t_1)$.
This defines the characteristic wavenumber of modes frozen as
$\Psi$ is passing through unity, 
\beq
	k_1/a_1=\mu\epsilon ^{-1/2}.\label{eq:k1}
\label{eq:22}
\eeq
A mode with wavenumber $\sim k_1$ is frozen at the value
$\phi _{k_1}^o\sim 1/a_1$ (eq~[\ref{eq:bdy}]), so the rms field
fluctuation on this scale is frozen at the value 
\beq
 	\phi _1^2\sim k_1{}^3P_\phi (k_1)\sim
	(k_1/a_1)^2\sim H_1{}^2. 
\label{eq:23}
\eeq
Because the field has become effectively massless it is frozen at
a value on the order of the Hubble parameter at freezing. At shorter
wavelengths, the modes are frozen later, at smaller $H$. This is
the wanted convergence of the field fluctuations on small scales.  

At wavelengths much longer than $k_1^{-1}$ the mode stops
oscillating at $a_k\ll a_1$. In the time interval 
$a_k \ll a\ll a_1$ the wavenumber in the second part of
equation~(\ref{eq:fieldeqs}) is unimportant and the Higgs field
varies as $\Psi\propto 1/t$ (eq.~[\ref{eq:psiearly}]). The result
is that the mode amplitude varies as a power law of the time, 
\beqa
	\phi _k\propto a^{-q}, \qquad\qquad\qquad\nonumber\\
	2q = 3 - \epsilon -[(3-\epsilon )^2 
	- 4\epsilon (3-2\epsilon )\gamma /\beta ]^{1/2}.
		\label{eq:q}
\eeqa
This is the less rapidly decaying solution to the second of
equations~(\ref{eq:fieldeqs}) when the term $\propto k^2$ may be
neglected. Thus the amplitude of
a long wavelength mode is $\phi _k\sim 1/a_k$ at Hubble  
crossing (eq.~[\ref{eq:bdy}]), and the amplitude decays by the 
factor $(a_k/a_1)^q$ from Hubble crossing to $\tau =1$. After
this $\Psi$ decreases more rapidly than $H$
(eq.~[\ref{eq:psilate}]) so the 
amplitude approaches the limiting value
\beq
	\phi _k^o\sim a_k^{q - 1}/a_1^q\propto
	k^{-(1 - q)/(1 - \epsilon )}.
\eeq
With equations~(\ref{eq:pphi}) and~(\ref{eq:q}), the 
power spectrum is
\beqa
	P_\phi\propto k^{m_\phi},\qquad\qquad\qquad\nonumber\\
	m_\phi = -[(3-\epsilon )^2 - 4\epsilon (3-2\epsilon )\gamma 
	/\beta ]^{1/2}/(1-\epsilon ), \label{eq:mphi}
\eeqa
on scales much larger than the
break at $k\sim\mu a_1\epsilon ^{-1/2}$ (eq.~[\ref{eq:k1}]).

Figure 1 shows an example of numerical solutions for the
evolution of the amplitudes of Fourier modes of the CDM
field. In the computation the amplitude is written as
\beq
	\phi _k(t)=A(t)\exp -i\int ^t\omega d\tau , 
\eeq
as in equation~(\ref{eq:bdy}), and the field equation is written as
a differential equation for $A(t)$. The numerical computation of
$A(t)$ is straightforward. The figure shows that 
$|\phi _k|^2$ decays as $1/a(t)^2$ until the
wavelength reaches the Hubble length (eq.~[\ref{eq:bdy}]), then 
decays as $\sim a^{-2q}$ until $\Psi$ starts to decrease rapidly
at $\tau\sim 1$ (eqs.~[\ref{eq:psilate}] and~[{\ref{eq:q}]), and
then approaches a constant value, as expected. The mode starts
oscillating again after inflation when the Hubble parameter has
decreased to  $H\sim m$, where $m$ is the final CDM mass. This is 
discussed in the next section. 

Figure~2 shows the power law index 
\beq
	m_e = d\log P_\phi/d\log k, \label{eq:mnum}
\eeq
for the power spectrum~(eq.~[\ref{eq:pphi}]) from the
numerical solution in Figure~1. The index approaches 
equation~(\ref{eq:mphi}) at $k\ll k_1$, and falls below the
scale-invariant value $m_e = -3$ at $k\gg k_1$, again as
expected. With the numerical values in the next section, 
wavelengths of interest for structure formation are some six
orders of magnitude longer than at the break in the spectrum, and
$P_\phi$ is quite close to a power law with the index $m_\phi$ in
equation~(\ref{eq:mphi}).

The primeval CDM mass distribution is proportional to the square
of $\phi ({\bf x})$ (eq.~[\ref{eq:rhophi}]). Since 
$\phi ({\bf x})$ is a Gaussian process with zero mean value, we
have 
\beq
	\langle\phi ({\bf x}_1)^2\phi ({\bf x}_2)^2\rangle =
	\langle\phi ^2\rangle + 
	2\langle\phi ({\rm x}_1)\phi ({\rm x}_2)\rangle ^2,
\eeq
so the dimensionless mass autocorrelation function is
\beq
	\xi _\rho (x_{12}) = 
	{\langle\rho ({\rm x_1})\rho ({\rm x_2})\rangle\over
	\langle\rho\rangle ^2} - 1  =
	2{\langle\phi ({\rm x}_1)\phi ({\rm x}_2)\rangle ^2\over
	\langle\phi^2\rangle ^2}.\label{eq:xirho}
\eeq
That is, the mass autocorrelation function is proportional to the 
square of the field autocorrelation function. If the
power spectrum of the field fluctuations is 
$P_\phi\propto k^{m_\phi}$ (eq.~[\ref{eq:mphi}]), with
$-3/2>m_\phi >-3$, then the Fourier transform is 
$\xi _\phi\propto r^{-(3+m_\phi )}$, the mass autocorrelation
function is 
$\xi _\rho\propto\xi _\phi ^2\propto r^{-(6+2m_\phi )}$, 
and the Fourier transform is the mass fluctuation power spectrum
\beq
	P_\rho\propto k^{m_\rho }, \qquad
	m_\rho = 3 + 2m_\phi .\label{eq:mrho}
\eeq
Equation~(\ref{eq:xirho}) indicates the small-scale
primeval mass fluctuations are mildly nonlinear,
$\xi _\rho (0) = 2$ or $\delta\rho /\rho = 2^{1/2}$.

The fit to the second moments of the large-scale mass and CBR
distributions requires $m_\rho\sim -1.8$ (Paper~II).
The integral over this power law diverges at 
large $k$, so there has to be a break from the power law 
shortward of which $P_\rho$ decreases more rapidly than $k^{-3}$. 
This break is produced by the mass $\mu$
of the Higgs field, which defines the coherence length of the
frozen field fluctuations (eq.~[\ref{eq:k1}]),
\beq
	x_1\sim\mu ^{-1}\epsilon ^{1/2}a_1{}^{-1}.
		\label{eq:x1}
\eeq
After inflation and at $H\gg m$ the mass density in 
the CDM is determined by the rms value of the field frozen at 
$a\sim a_1$ (eq~[\ref{eq:23}]), 
\beq
	\rho _\phi = m^2\langle\phi ^2\rangle /2 \sim 
	m^2\phi _1{}^2 \sim m^2H_1{}^2.\label{eq:phi1}
\eeq

\section{Numerical Values}

This section gives numerical values for the parameters in the
inflation model based on a fit to the observational constraints.
The computations are carried to fractional accuracy of first order
in $\epsilon$. 

For definiteness I adopt the cosmological model parameters
\beqa
	\Omega = 0.2, \qquad\Omega _{\rm B}\lsim 0.05,
	\qquad \lambda =0.8, \nonumber\\
	T_o=2.73\hbox{ K},\qquad
	H_o=70\hbox{ km s}^{-1}\hbox{ Mpc}^{-1},
\label{eq:cosmologicalparameters}
\eeqa
and the inflation model parameters
\beq
	\epsilon = 0.1,\ \gamma /\beta = 3.3, \	m_\phi = -2.4,
	\ m_\rho = -1.8. 
\label{eq:parameters}
\eeq
The three families of neutrinos are massless. 
The model universe is cosmologically flat; $\lambda$
is the contribution of a term in the stress-energy tensor
that acts like Einstein's cosmological constant. The density
parameter in baryons is $\Omega _{\rm B}$, and the density
parameter in the CDM is $\Omega -\Omega _{\rm B}$. The density
parameter is reduced from values considered in Peebles~(1997) to
get a better cluster 
mass function, following the advice of Cen (1997). This is 
discussed further in Paper~II. The first two parameters in
equation~(\ref{eq:parameters}) produce the tilted power law
index $m_\phi$ for the field power spectrum, which translates to 
the mass fluctuation spectrum power law index $m _\rho = -1.8$ 
(eqs.~[\ref{eq:mphi}] and~[\ref{eq:mrho}]). This gives a
reasonable fit to the spectrum of angular fluctuations of the 
thermal cosmic background radiation and the large-scale
fluctuations of the present galaxy distribution, which are
assumed to be traced by the galaxies (Peebles~1997; paper~II).
The same value of $m_\phi$ follows from a smaller value of
$\epsilon$ and a larger value for $\gamma /\beta$, a case that is
not considered here. 

\subsection{Parameters for Inflation}

Here I review common measures of the expansion
factor during and after inflation. If the temperature at the end of
inflation is $T_r$, and the entropy is shared among
about as many particles and spin states as now, the expansion
factor from the end of inflation to the present is
\beq
	{a_o\over a_r}\sim {T_r\over T_o}\sim 10^{27}T_{14},
		\label{eq:ao/ar}
\eeq
where $T_o=2.73$~K and $T_r$ is expressed in units of
$10^{14}$~GeV. The mass density $\sim aT_r^4$ fixes the Hubble
parameter $H_r$; the ratio to the present value in
equation~(\ref{eq:parameters}) is
\beq
	H_r/H_o \sim 10^{51}T_{14}^2. \label{eq:Hr/Ho}
\eeq
Fluctuations in $\phi$ on the scale of the present Hubble length
$H_o^{-1}$ are frozen at epoch $a\sim a_p$ when the Hubble length
is $H_p^{-1}$. The expansion factor from then to the present is  
\beq
	a_o/a_p = H_p/H_o. 
\eeq
If the power law expansion $a\propto t^{1/\epsilon }$ persists
to the end of inflation then the expansion factor from $a_p$ to
$a_r$ is
\beq
	{a_r\over a_p}\sim 
 \left( {a_r\over a_o}{H_r\over H_o}\right) ^{1/(1-\epsilon)}
	\sim 10^{27}T_{14}{}^{1.1}, \label{eq:ar/ap}
\eeq
for $\epsilon = 0.1$. This is in the familiar range of values
(Kolb \&\ Turner 1990).  

\subsection{Adopted Inequalities}

The solutions for $\psi$ and the CDM field $\phi$ in 
\S 2 are approximations based on inequalities to be discussed
here. The condition  
\beq
	\gamma\phi ^2\ll\mu ^2\lsim \beta\psi ^2,\label{eq:bound}
\eeq
at $\tau\lsim 1$ assures that the energy density in $\phi$ is
small compared to that in $\psi$, and that the coupling to $\phi$ 
has little effect on the evolution of $\psi$, as assumed in \S 2.
I adopt $\epsilon =0.1$, not a very small number, so the
wanted power law index $m_\phi$ requires  
$\gamma\sim\beta$. Since $\phi$ is frozen at a value comparable to
$H=1/\epsilon t$, equations~(\ref{eq:psiearly}) and~(\ref{eq:bound})
require
\beq
	\epsilon\gg\gamma\sim\beta .\label{eq:betabound}
\eeq	

The computation assumes the energy density is dominated by 
that of the field $\chi$ that is driving inflation 
(eq.~[\ref{eq:rho1}]). At $\tau\lsim 1$ this condition is
\beq
	\rho _\psi \simeq {9\over 4\beta\epsilon ^2t^4} \ll 
	{3m_{\rm pl}^2\over 8\pi\epsilon ^2t^2}, 
	\quad (m_{\rm pl}t)^2\gg 6\pi /\beta .\label{eq:rhopsi}
\eeq
The above numbers imply that the time $t_p$ of freezing of the
fluctuations we observe at the present Hubble length satisfies
\beqa
	m_{\rm pl}t_p&\sim &{10^{9.8 - 24\epsilon}\over
	\epsilon T_{14}^{2+\epsilon}}\nonumber\\
	&\sim &10^8T_{14}{}^{-2.1}
	\gg (6\pi\beta ^{-1})^{1/2},
\eeqa
for $\epsilon =0.1$. Under the condition to be discussed next,
that inflation is cool enough that curvature fluctuations are
subdominant, this condition applied at the time the density
fluctuations we can observe are produced is easily reconciled
with the condition that $\beta\ll\epsilon$ (eq.~[\ref{eq:betabound}]).

\subsection{Primeval Adiabatic Density Fluctuations}

The isocurvature model assumes primeval curvature (adiabatic)
fluctuations are subdominant on scales of interest for structure
formation. I consider first the effect of the density
fluctuations frozen into the CDM field $\phi$, and then those of
the field $\chi$ that drives inflation. 

In time-orthogonal coordinates the components of the metric
tensor are $g_{00}=1$, $g_{0\alpha}=0$, and
\beq
	g_{\alpha\beta} = -a(t)^2\left( \delta _{\alpha\beta} 
	- h_{\alpha\beta} \right).
\eeq
In linear perturbation theory the space-time part of the Einstein
field equation is (Peebles 1980)
\beq 
	\onehalf (\dot h_{,\alpha} -\dot h_{\alpha\beta ,\beta})
	= 8\pi GT_{0\alpha}\sim \dot\phi\phi _{,\alpha}/m_{\rm pl}{}^2.
		\label{eq:T_0alpha}
\eeq
To see the meaning of the combination of terms in the
perturbation $h_{\alpha\beta}$ to spacetime curvature 
consider the Fourier transform of the field equation.
With coordinates oriented so the wave vector {\bf k} of a Fourier
component is parallel to the $x^3$ axis,
equation~(\ref{eq:T_0alpha}) is 
$ik\dot h_{11}=8\pi GT_{03}({\bf k})$. The component $h_{33}$ along the
direction of {\bf k} does not appear in this equation and the
value is of no interest because it is freely adjusted by a 
coordinate relabeling along the $x^3$ axis. The transverse part
$h_{11}=h_{22}$ for the inhomogeneous solution gives a measure of 
the curvature fluctuation produced by the freezing of the field
component,
\beq
	h_{11}=-8\pi i G\int T_{03}/k\, dt\sim \delta (t)[a(t)/kt]^2.
\label{eq:h11}
\eeq
As discussed in Peebles (1980, \S 86), and
indicated in the last expression, $h_{11}$ determines the
amplitude of the mass density contrast  
$\delta =\delta\rho /\rho$ in the most rapidly growing adiabatic mode 
at the end of inflation. When the proper wavelength again approaches
the Hubble length, at $a/k\sim t$, the 
observed adiabatic density contrast on the scale of the Hubble
length is $(\delta\rho /\rho )_h\sim h_{11}$. I keep the notation
$h_{11}$ for the result of summing over Fourier components to get
the curvature fluctuation frozen on the comoving scale $k^{-1}$, 
which is equal to the adiabatic CBR temperature fluctuation 
$\delta T/T\sim h_{11}$ appearing at the Hubble length. 

As fluctuations in the CDM field $\phi$ are freezing the factors
$\dot\phi$ and $\phi _{,\alpha}/a$ both are on the order of 
$H\phi\sim H^2$, so equation~(\ref{eq:T_0alpha}) indicates 
the temperature fluctuation appearing now at the Hubble length
is
\beq
	h_{11}\sim (H_p/m_{\rm pl})^2\sim 
	10^{-20+54\epsilon }T_{14}{}^4 \ll 10^{-5}.\label{eq:h11p}
\eeq
The last expression is the condition that the
adiabatic perturbation resulting from the freezing 
of $\phi$ is small compared to the isocurvature perturbation that
will be tuned to match the observations. The same condition
eliminates the tensor quantum fluctuations frozen into the
spacetime curvature.  

The adiabatic perturbation from the fluctuations frozen into the
field that is driving the power law inflation produces CBR
temperature anisotropy
\beq
	\delta T/T\sim (T_r/m_{\rm pl})^2\ll 10^{-5}.
\label{eq:eq48}
\eeq
At $\epsilon =0.1$ and $T_{14}\lsim 1$ the conditions in 
equations~(\ref{eq:h11p}) and~(\ref{eq:eq48}) are well satisfied;
curvature fluctuations are negligibly small.

\subsection{The Coherence Length of the CDM Field}

The parameters for the inflation model are constrained by the
coherence length of $\phi ({\bf x})$ at the break in its power
law spectrum (eq.~[\ref{eq:k1}]): the coherence length
fixes the mean square value of the field, which in turn fixes the
mean mass density in CDM, $\rho =m^2\langle\phi ^2\rangle$.

The required value of the coherence length follows from the
normalization to the mass fluctuation spectrum. The spectrum
extrapolated to the present in linear perturbation theory is
taken to be  
\beqa
	P_\rho (k) &=& \int d^3r\,\xi_\rho e^{i{\bf k}\cdot {\bf r}}
		\nonumber\\
	&=& 6300h^{-3}(0.1h/k)^{1.8}\hbox{ Mpc}^3.
\label{eq:prho} 
\eeqa
This is consistent with the assumption that galaxies trace mass
on large scales (Paper~II). The unit for the wavenumber is
radians per megaparsec, and the power law index is the adopted
value in equation~(\ref{eq:parameters}). 
The Fourier transform of $P_\rho$ is the mass autocorrelation
function in linear perturbation theory,
\beq
	\xi _\rho = 7.2(hr_{\rm Mpc})^{-1.2}.
\label{eq:xigg}
\eeq

The primeval mass distribution is homogeneous, with fluctuations
in the local CDM mass density balanced by opposite fluctuations
in the baryons and radiation. As these fluctuations in
composition appear at the Hubble length the radiation 
distribution is smoothed, at short wavelengths by the propagation
of acoustic waves that are dissipated by photon diffusion in the
baryon-radiation fluid before decoupling, and at large
wavelengths by free streaming of the radiation after decoupling. 
This leaves the CDM as an isothermal distribution within the
Hubble length. Wavelengths in the CDM distribution that
reach the Hubble length  prior to the epoch $z_{\rm eq}$ of
equality of mass densities of matter and radiation are left with close to
the primeval amplitude at $z = z_{\rm eq}$. At longer wavelengths
the CDM distribution at $z_{\rm eq}$ is smoother than the
initial value because the net mass density
remains close to homogeneous when pressure gradient forces may be
neglected. 

At wavelengths small compared to the Hubble length at 
$z=z_{\rm eq}$ the amplitude of a CDM fluctuation varies with
time as the isothermal perturbation equation 
\beq
	\ddot D + {2\dot a\over a}\dot D =
	4\pi GD\rho _{\rm CDM} = 
	{3\over 2}\left(a_o\over a\right) ^3\Omega DH_o{}^2,
\label{eq:deltac}
\eeq
where the expansion rate equation is
\beq
	(\dot a/a) ^2 = H_o{}^2
	\left[\Omega\left( a_o/a\right) ^3 +
	\Omega _r\left( a_o/a\right) ^4 + \lambda\right] .
\label{eq:adot}
\eeq
The density parameter in radiation is
\beq
	\Omega _r = 4.2\times 10^{-5}h^{-2}.
\eeq
The density parameter in baryons plus CDM is $\Omega$, and
equation~(\ref{eq:deltac}) assumes the CDM mass density is much
larger than that of the baryons. A more complete analysis would
describe the collapse of 
the baryons into the CDM mass concentrations after decoupling,
but the resulting correction to equation~(\ref{eq:deltac}) is
beyond the level of accuracy of this discussion. 
Equations~(\ref{eq:deltac}) and~(\ref{eq:adot}) assume a
cosmologically flat model, with
$\lambda = 1-\Omega -\Omega _r$. 

Results of numerical integration of these 
equations with $h=0.7$ are shown in Figure~3. The net growth
factor from high redshift to the present is well
approximated by the fitting function 
\beq
	e(\Omega ) = D(z\rightarrow\infty )/D(z=0) = 
	5.7\times 10^{-5}\Omega ^{-1.22}.
\label{eq:do}
\eeq
In linear perturbation theory the primeval CDM mass
autocorrelation function is $e(\Omega )^2$ times 
the present value in equation~(\ref{eq:xigg}) at fixed comoving
length. The field autocorrelation function is related to the
mass function by equation~(\ref{eq:xirho}). The result is the
primeval field autocorrelation function,
\beq
	{\langle\phi _1\phi _2\rangle\over\langle\phi ^2\rangle}
	= {1\times 10^{-4}\over (hx_{\rm Mpc})^{0.6}\Omega ^{1.22}}.
\eeq
At $\Omega = 0.2$ this power law extrapolates to unity at
\beq
	hx_1 = 7\hbox{ pc}.\label{eq:x1p}
\eeq
In the approximation that the shape of the power spectrum
changes fairly rapidly from $P_\rho(k)\propto k^{-1.8}$ 
on large scales to $P_\rho\propto k^s$ with $s<-3$ on smaller
scales, equation~(\ref{eq:x1p}) is the required comoving coherence
length of the CDM field $\phi (\bf x)$.

\subsection{The CDM Mass and Mass Density}

Now let us consider the condition that the energy density in
$\phi$ left from inflation matches the present mean mass density
in the cosmological model.

The present cosmological mass density is 
\beqa
 	\rho _o &=& 1.88\times 10^{-29}\Omega h^2\hbox{ g cm}^{-3}\\
	&=& 8\times 10^{-47}\Omega h^2\hbox{ GeV}^4\\
	&=& m^2\phi _o{}^2,\label{eq:phio}	
\eeqa
where $m$ is the field mass and $\phi _o$ is the present rms
value of the field. This gives 
\beq
	\phi _o=10^{-23}h\Omega ^{1/2}m_9{}^{-1}\hbox{ GeV},
\eeq
where
\beq
	m = m_9\hbox{ Gev}.
\eeq
The field fluctuations are nonrelativistic, because the de~Broglie
wavelength at the coherence length $a(t)x_1$ is much longer than the
Compton wavelength all the way back to inflation, so at $H<m$
the rms field value varies as $\phi\propto a(t)^{-3/2}$. When the
universe is radiation-dominated the Hubble parameter is 
\beqa
	H&\sim & 2\times 10^{-20}(a_o/a)^2\hbox{ s}^{-1}\nonumber\\ 
	&=&1\times 10^{-44}(a_o/a)^2\hbox{ GeV},
\eeqa
if the number of particle families and spins at very high
redshift is about the same as at $z\sim 10^{10}$. The field
starts oscillating after inflation at $H\sim m$, when the
expansion factor is
\beq
	a_o/a_m\sim 1\times 10^{22}m_9{}^{1/2}.
\eeq
The field thus is frozen at the rms value
\beqa
	H_1&\sim &\phi _1\sim\phi _o(a_o/a_m)^{3/2}\nonumber\\
	&\sim &	7\times 10^9m_9{}^{-1/4}h\Omega ^{1/2} \label{eq:phif}\\
	&\sim & 2\times 10^9m_9{}^{-1/4}\hbox{ GeV},\nonumber
\eeqa
for the cosmological parameters in
equation~(\ref{eq:cosmologicalparameters}). 

Equation~(\ref{eq:phif}) fixes the value of the Hubble parameter
when the field coherence length is equal to the Hubble length 
(eq.~[\ref{eq:23}]). The model assumes power law expansion, 
$H\propto a^{-\epsilon}$, continues to the
end of inflation at expansion parameter $a_r$, so
the present value of the coherence length is 
\beq
	r_1=H_1{}^{-1}{a_o\over a_1}\sim 
	H_r{}^{-1}\left(a_r\over a_1\right) ^{1-\epsilon }
	\left(a_o\over a_r\right).
\eeq
With equations~(\ref{eq:ao/ar}) and~(\ref{eq:Hr/Ho}) and
$\epsilon =0.1$ this gives the expansion factor from freezing of
the field at its coherence length to the end of inflation,
\beq
	a_r/a_1	\sim 10^{16}[r_1({\rm pc})T_{14}]^{1.1}.
		\label{eq:ar/a1}
\eeq
This is much smaller than the expansion factor at freezing of the
field fluctuations on the scale of the present Hubble length, of
course (eq~[\ref{eq:ar/ap}]). 

The rms value of the frozen field is
$\phi _1\sim H_1\sim H_r(a_r/a_1)^\epsilon$. The temperature at
the end of inflation follows from
equations~(\ref{eq:Hr/Ho}), (\ref{eq:phif}),
and~(\ref{eq:ar/a1}):
\beq
	T_{14}\sim 0.2 m_9^{-0.12}r_1{(\rm pc})^{-0.05}.
\eeq
This indicates that for a considerable range of 
values of the CDM particle mass, $m$, inflation would have to end
at a relatively cool temperature, about $10^{13}$~GeV.  
Finally, the mass $\mu$ in the Lagrangian~(\ref{eq:L}) is given
by equations~(\ref{eq:x1}) and~(\ref{eq:phif}),
\beq
	\mu\sim\epsilon ^{1/2}H_1\sim 10^9m_9{}^{-1/4}
		\hbox{ GeV}\label{eq:mu}
\eeq

\section{Discussion}

In the adiabatic CDM family of models (that I have termed ACDM)
the cold dark matter particles can have been
produced out of entropy at any time after inflation ended, a
broadly general situation. The isocurvature CDM (ICDM) model is
more specific --- the CDM is a remnant of a field present
during inflation --- and hence arguably less likely in the
absence of evidence for or against the nature of the CDM. This
consideration may become more compelling when we have better
experimental constraints on the dark matter. 

Assessments of the relative degree of ``fine tuning'' of the
adiabatic and isocurvature inflation models also seem to be of
doubtful significance at this stage of our understanding, so I
note only four aspects of the ICDM model. First, there is
precedent in the literature for the CDM potential  $V(\psi ,\phi
)$ in equation~(\ref{eq:V}). The same functional form, applied in
a different context, appears in the Hybrid Inflation model
(Copeland et al. 1994; Linde 1994), and similar forms are not
uncommon (eg. Kofman \& Linde 1987; Hodges et al. 1990; Randall,
Solja\u ci\'c \& Guth 1996). Second, the conditions on an
acceptable set of dimensionless parameters for the ICDM model are
$\gamma\sim\beta\ll\epsilon\sim 0.1$. These do not seem
unduly severe or artificial. Third, the model requires that the
initial value of the field $\psi$ be large enough that $\psi$
approaches the solution in  equation~(\ref{eq:psiearly}) well
before field fluctuations are frozen on scales we can observe,
and that the initial value of $\phi$ be small enough that $\psi$
can drive it to zero mean value before the field is squeezed on
scales of interest. I do not know how to judge whether these
conditions are likely outcomes of the physical situation prior
to inflation. Finally, an acceptable set of values of the
characteristic masses that appear in the model is 
\beqa	
M \sim 10^{19}\hbox{ GeV},\qquad 
T_r \sim  10^{13}\hbox{ GeV},\nonumber\\
\mu \sim  10^{9}\hbox{ GeV},\qquad	
m \sim  1 \hbox{ GeV}. 
\eeqa 
The broad range of values is
impressive, but so is the range of measured characteristic masses
in particle physics and those that appear in physics-motivated
models for inflation (Randall 1997). 

The greatly tightened constraints from observational programs in
progress are going to make it much harder to invent acceptable
structure formation models such as ACDM and ICDM from largely
theoretical and aesthetic considerations. Some already known member of
the CDM family --- in which I would include ACDM and ICDM --- may
survive the precision tests in progress and by its success compel
acceptance as a truly valid approximation to reality. If this
does not happen, perhaps we will be lucky enough to see in the
results from observations and particle physics
guidance to the formulation of more promising models.

I conclude that the ICDM model has as valid a pedigree from
accepted ideas about the very early universe as might be expected
for a model that is motivated by astronomical phenomenology
rather than physics. The observational situation is discussed in
Paper~II.

\acknowledgements

This work was supported in part by the NSF.

\plotone{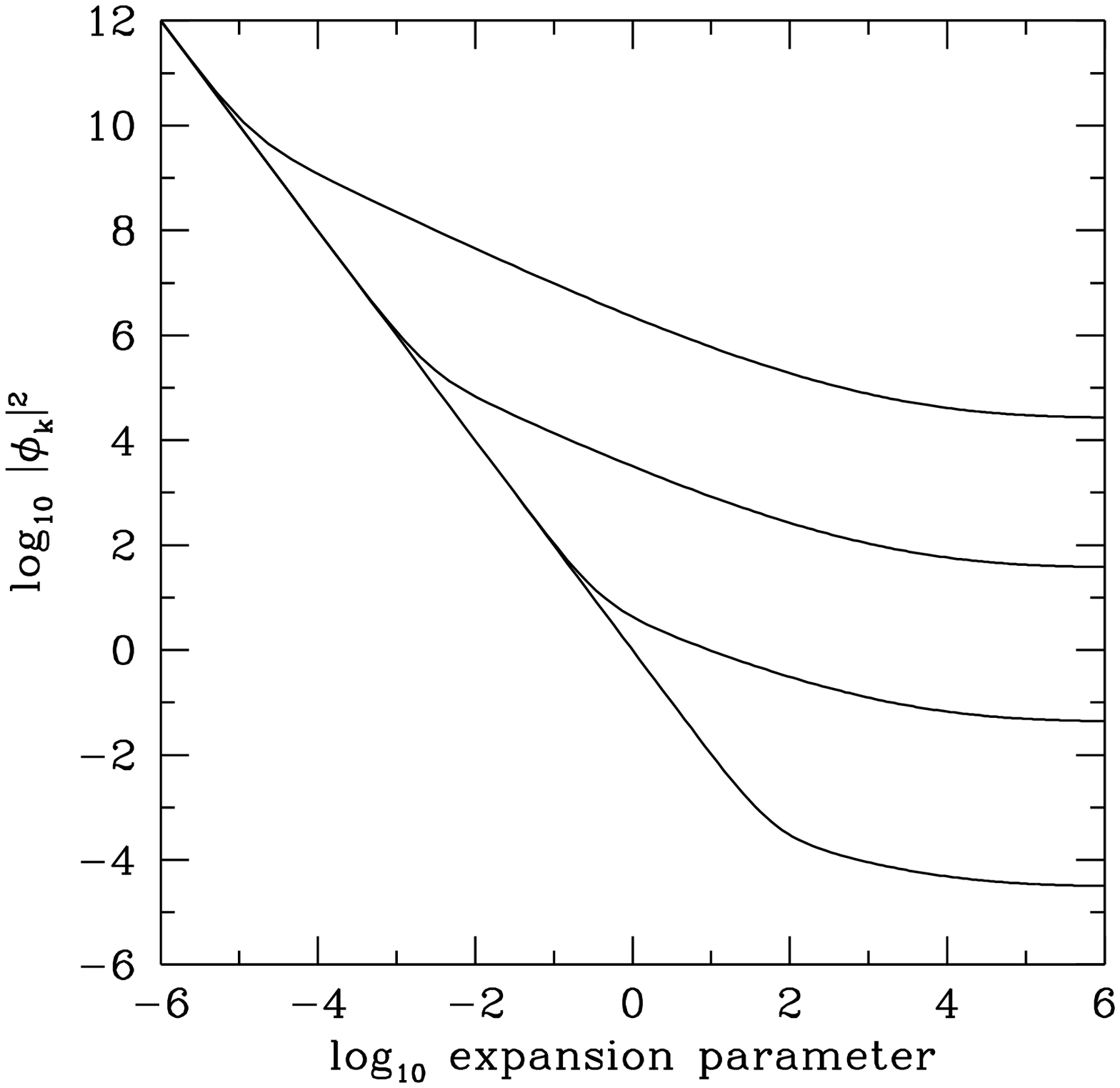}
\figcaption[figure1I.ps]{Evolution of amplitudes of Fourier modes
of the CDM field for $\epsilon =0.1$ and $\gamma =3.3\beta$. The
dimensionless wavenumbers are $\kappa = 10^{-4}$, $10^{-2}$, 1,
and 100 from top to bottom.  \label{fig1}}

\plotone{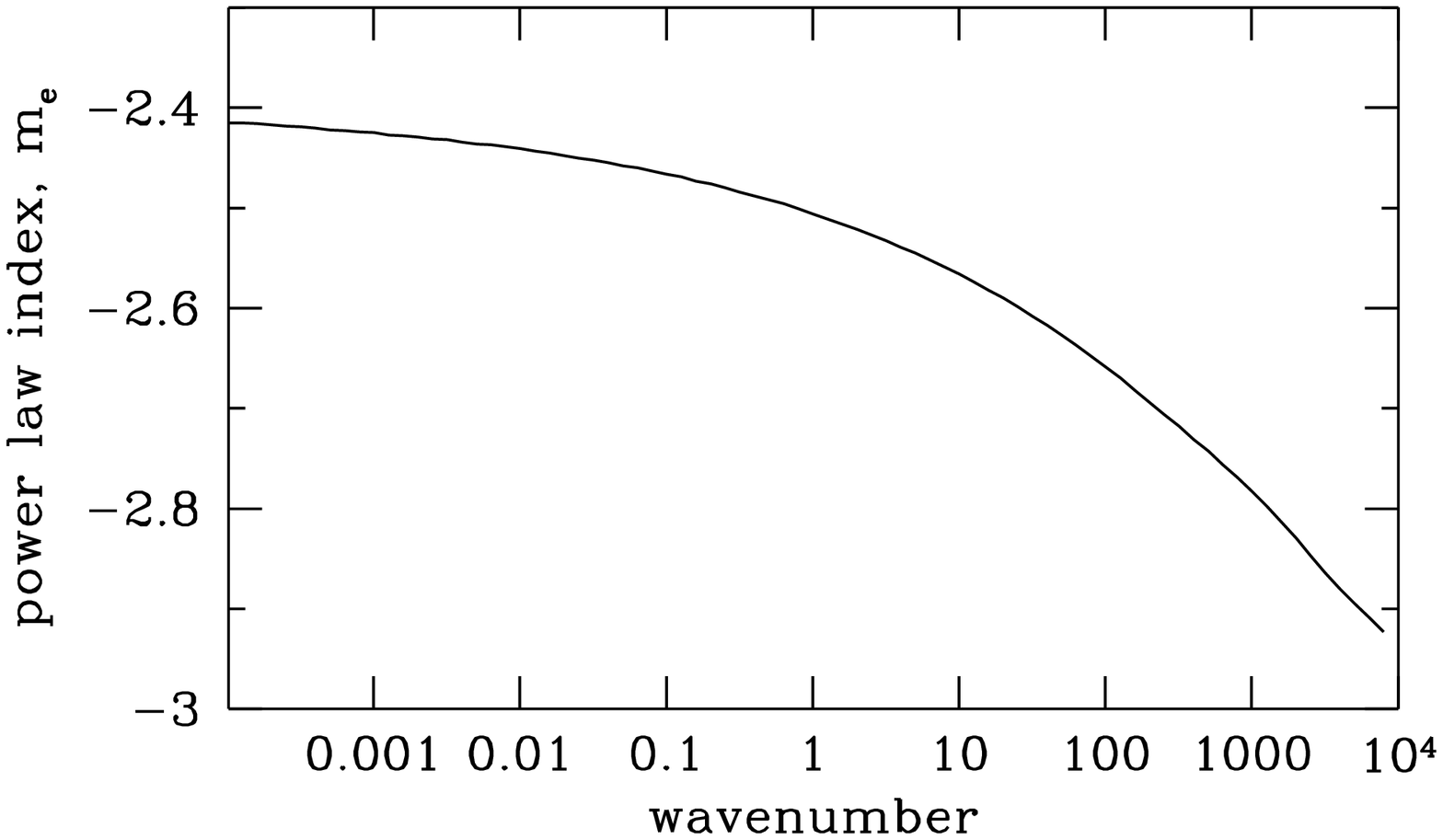}

\figcaption[figure2I.ps]{Logarithmic derivative of the CDM field
power spectrum (eq.~[\ref{eq:mnum}]) for a numerical solution
with the parameters in Figure~\ref{fig1}. This shows the region
near the break in the spectrum.  \label{fig2}}

\plotone{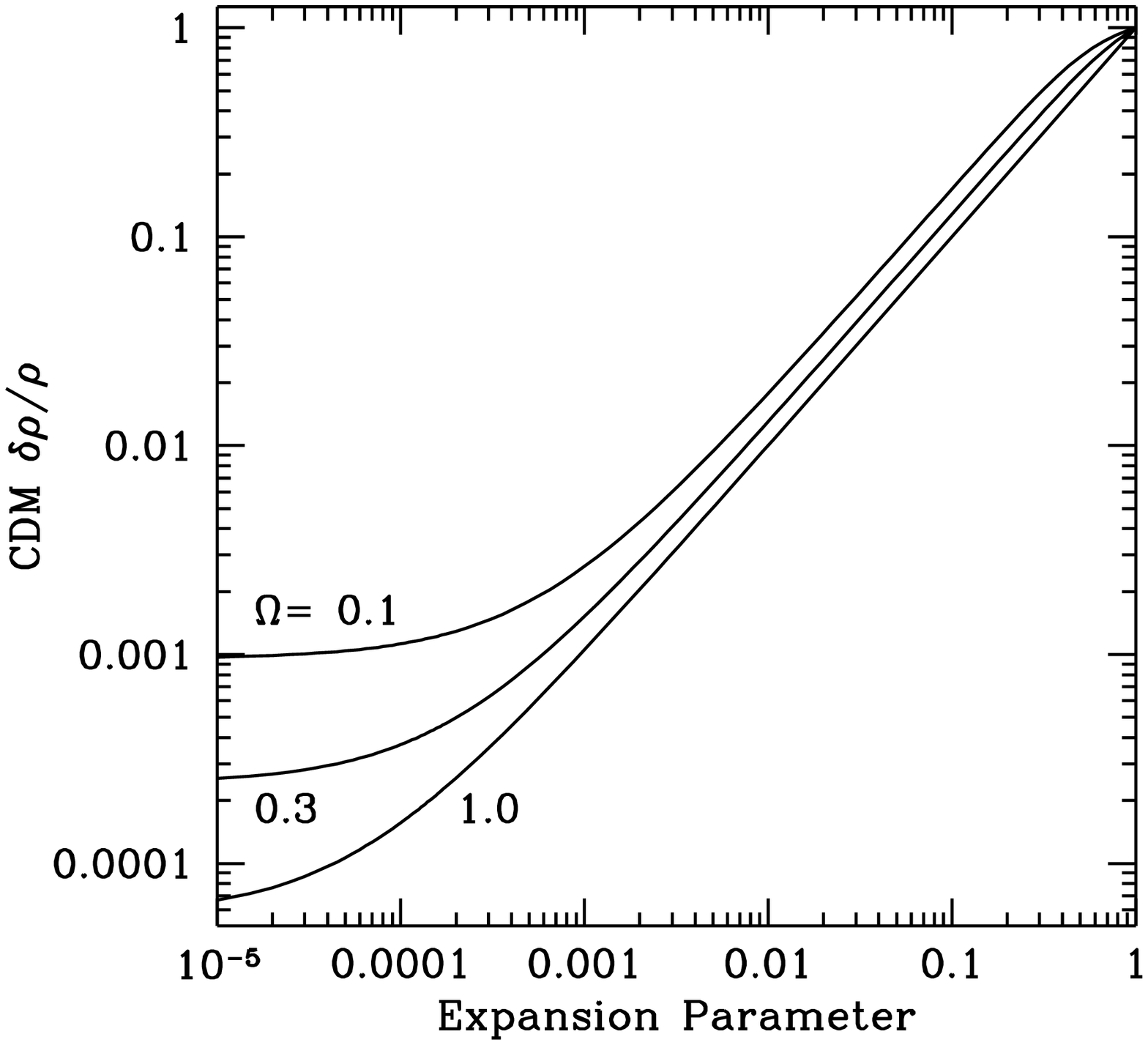}

\figcaption[figure3I.ps]{Evolution of the CDM density contrast on
scales small compared to the Hubble length at redshift 
$z_{\rm eq}$, in linear perturbation theory. The primeval
isocurvature density fluctuation 
becomes an isothermal CDM fluctuation when its length scale
becomes comparable to the Hubble length and
radiation pressure smooths the distribution of radiation and 
baryons. The isothermal density fluctuation starts to grow as the 
redshift approaches $z_{\rm eq}$. \label{fig3}}

\end{document}